\documentclass[]{interact}

\usepackage{epstopdf}% To incorporate .eps illustrations using PDFLaTeX, etc.
\usepackage[caption=false]{subfig}% Support for small, `sub' figures and tables

\usepackage{microtype}                 % use micro-typography (slightly more compact, better to read)
\PassOptionsToPackage{warn}{textcomp}  % to address font issues with \textrightarrow
\usepackage{textcomp}                  % use better special symbols
\usepackage{mathptmx}                  % use matching math font
\usepackage{times}                     % we use Times as the main font
\usepackage{tabu}                      % only used for the table example
\usepackage{booktabs}                  % only used for the table example
\usepackage{amsmath,amsopn,amssymb}
\usepackage[ruled,linesnumbered,commentsnumbered,inoutnumbered,noend]{algorithm2e}

\SetCommentSty{customcommentsfont}

\usepackage{listings}
\usepackage{enumitem}
\usepackage{multirow}
\usepackage{tabularx}
\usepackage{xspace}
\usepackage{color}
\usepackage{soul}
\usepackage{graphicx}

\newcommand{\app}{Exploropleth\xspace}

\definecolor{orange}{RGB}{237,125,49}
\definecolor{purple}{RGB}{112,47,160}
\definecolor{pink}{RGB}{245,114,255}
\definecolor{black}{RGB}{0,0,0}
\definecolor{red}{RGB}{231, 76, 60}
\definecolor{teal}{RGB}{0, 128, 128}
\definecolor{gray}{RGB}{128, 128, 128}
\definecolor{cmaroon}{RGB}{192,0,0}
\definecolor{cbrown}{RGB}{197,90,17}
\definecolor{cpurple}{RGB}{112,48,159}
\definecolor{cgreen}{RGB}{84,130,53}
\definecolor{cblue}{RGB}{0,112,192}
\definecolor{clightgray}{RGB}{230, 230, 230}

\newcommand{\texticon}[1]{\includegraphics[height=1\fontcharht\font`\B]{src/camera-ready-figures-tables/icons/#1.eps}}
\newcommand{\expert}[1]{\textcolor{teal}{$E_{#1}$}}

\newcommand{\paragraphHeadingSpace}{\vspace{8px}}
\newcommand{\bpstart}[1]{
\noindent{\textbf{#1}}%
}

\usepackage[natbibapa,nodoi]{apacite}% Citation support using apacite.sty. Commands using natbib.sty MUST be deactivated first!
\theoremstyle{plain}% Theorem-like structures provided by amsthm.sty

\theoremstyle{definition}

\theoremstyle{remark}

\begin{document}

% \articletype{Research}%

\title{Exploropleth: Exploratory Analysis of Data Binning Methods in Choropleth Maps}

% Authors
\author{
\name{Arpit Narechania\textsuperscript{a,b}\thanks{Emails: arpit@ust.hk, endert@gatech.edu, clio@gatech.edu. Corresponding Author: Arpit Narechania}, Alex Endert\textsuperscript{b} and Clio Andris\textsuperscript{b}}
\affil{\textsuperscript{a}The Hong Kong University of Science and Technology, Hong Kong SAR, China}
\affil{\textsuperscript{b}Georgia Institute of Technology, Georgia, USA}
}

\maketitle

% Abstract
\begin{abstract}
When creating choropleth maps, mapmakers often bin (i.e., group, classify) quantitative data values into groups to help show that certain areas fall within a similar range of values. For instance, a mapmaker may divide counties into groups of high, middle, and low life expectancy (measured in years). It is well known that different binning methods (e.g., natural breaks, quantile) yield different groupings, meaning the same data can be presented differently depending on how it is divided into bins. To help guide a wide variety of users, we present a new, open source, web-based, geospatial visualization tool, \textbf{\app}, that lets users interact with a catalog of established data binning methods, and subsequently compare, customize, and export custom maps. This tool advances the state of the art by providing multiple binning methods in \textit{one view} and supporting administrative unit reclassification \textit{on-the-fly}. We interviewed 16 cartographers and geographic information systems (GIS) experts from 13 government organizations, non-government organizations (NGOs), and federal agencies who identified opportunities to integrate \app into their existing mapmaking workflow, and found that the tool has potential to educate students as well as mapmakers with varying levels of experience. \app is open-source and publicly available at \textbf{\url{https://exploropleth.github.io}}.
\end{abstract}

% Keywords

% Sections
\section{Introduction}
\label{section:introduction}

A choropleth map is a type of thematic map where administrative units or geographic regions are represented in different colors, shades, or patterns to represent the magnitude of an attribute associated with the underlying geography. 
A key part of creating choropleth maps is data binning (or classification), a process that transforms continuous data values into discrete bins (or classes)~\citep{de2007geospatial}. This process is widely used in statistical data analysis, pattern recognition, and machine learning and is essential for converting raw data into a visual representation that highlights spatial patterns--treating the map as a statistical surface~\citep{altman2006cost, de2007geospatial}. In particular, choropleth mapmakers define bins to highlight patterns and trends across geographic regions~\citep{de2007geospatial} and, in some cases, to preserve data privacy~\citep{lin2002using}. Map readers find these few, well-defined bins more manageable and easier to understand as they help simplify and clarify the main takeaway(s) from the map~\citep{sahann2021histogram}, despite the loss of absolute numeric data relationships~\citep{tobler1973choropleth}.

However, determining these bins is challenging as there is no single, ideal method. Cartographers and statisticians have developed and critiqued a suite of binning methods over the past 50 years. These methods use statistical indicators such as the data mean and standard deviation, spatial indicators such as the geographic area, or heuristics provided by subject matter experts~\citep{slocum2022thematic, andrienko2001choropleth, de2007geospatial}. Additional factors such as the distribution of data (e.g., is it normally distributed?), map purpose (e.g., can it highlight outliers?), mapmaker preferences  (e.g., is there a \emph{favorite, go to} method?) or organization styles  (e.g., is there an in-house color palette?), and audience (e.g., is it for domain experts or novices?) may also influence binning method choice~\citep{narechania2025cartographersincubicles}. 
Applying a binning method without careful consideration may create false patterns on the map misrepresent the actual geographic phenomena, which is both ineffective and misleading~\citep{monmonier2018lie}.

There are many popular software applications that help cartographers explore and apply different binning methods. For instance, libraries such as PySAL~\citep{pysal} provide an application programming interface (API) for customizing and integrating binning methods into users' own applications. Similarly, popular GUI-based tools such as ArcGIS~\citep{arcgis} and QGIS~\citep{qgis} let users visualize and interact with the binning outputs on a map. However, these popular applications sometimes omit newer or less common methods, and do not support client-light exploration. Nor do they facilitate comparison of methods in one view, as their focus is on wider, more computationally complex processes.

We believe the curation and communication of the wide established cartographic literature related to binning should have a facile entry point for both experienced and new mappers. Accordingly, we built a new geospatial visualization tool, \textbf{\app}, which shows visualizations of sixteen established binning methods as small multiples. Users can import their own data and interactively compare, customize, and export maps. This tool advances the state of the art by facilitating faceted browsing~\citep{yee2003faceted} of multiple binning methods in \emph{one view} and supporting administrative unit reclassification \emph{on-the-fly}. In doing so, \app helps convey the effects of different binning methods and shows that changing the binning method is not just a tweak, but can result in very different visualizations. Cartographers can use the tool to help explain how choosing a binning method has a sizable impact on the results of the map.   

We interviewed 16 cartographers and geographic information systems (GIS) experts from 13 government and non-government organizations (NGOs) and federal agencies to learn how \app can help them make better choropleth maps. These experts identified opportunities for integrating \app into their own mapmaking workflows, and highlighted its potential to educate expert as well as novice mapmakers and readers about different binning methods. They related that comparing different methods will encourage mapmakers to more thoughtfully choose an appropriate method, and learn to avoid unintentionally lying with maps.

Primary contributions of this work include:
\begin{enumerate}[nosep]
    \item An open source, web-based geospatial visualization tool, \textbf{\app},
    that helps users browse a catalog of established binning methods and interactively compare, customize, and export maps (Section~\ref{section:userinterface}). The tool can be publicly accessed at \textbf{\url{https://exploropleth.github.io}}.
    \item Findings from an evaluation with 16 cartographers and GIS experts on how \app can help choropleth mapmakers and readers (Section~\ref{section:expert-interviews}).
\end{enumerate}

We next outline related work, describe our design goals, demonstrate case studies, and outline the results of interviews. Finally, we discuss the implications and interpretations of our findings and conclude.
We also include a comprehensive overview of binning methods (including descriptions, references, examples, and usage guidelines) in Supplemental Material.

\section{Related Work}
\label{section:relatedwork}
\subsection{GIS Tools and Libraries}
Binning methods and approaches are built into GIS software, and reviewing the software helps us determine the availability, applicability, implementation feasibility, and popularity of different binning methods for integration into \app.

\paragraphHeadingSpace\bpstart{Tools.} 
In cartographic literature, EXPLOREMAP is a choropleth mapping tool to help mapmakers design maps, including the ability to reclassify the map data to determine the effect of different binning methods~\citep{exploremap1992egbert}.
GeoLinter is a linting framework for choropleth maps that, based on a set of design guidelines and metrics from cartographic literature, detects suboptimal mapping decisions and provides recommendations for improvement~\citep{lei2023geolinter}.
ChoroWare is a choropleth mapping tool to help mapmakers find ``best'' bin intervals; it uses evolutionary algorithms to generate solutions (bin intervals) for a multiobjective function enabling users to interactively explore them~\citep{xiao2006choroware}.
Descartes is a software system that supports visual exploration of spatially referenced data by automating map design and enabling interactive manipulation of maps for effective data analysis and problem-solving~\citep{andrienko1999interactive}.
Compared to these, \app{} is distinguished because it supports simultaneous browsing and comparison of multiple binning methods in a single view, allows for dynamic reclassification by directly manipulating administrative regions, and is web-based and open-source.

There also exist commercial tools that have enabled users to create, manage, and analyze geospatial data as choropleth maps for applications in public health, environmental science, tourism, and urban planning. Among these, ArcGIS~\citep{arcgis} is the most popular commercial tool, offering seven binning methods. QGIS~\citep{qgis} and GRASS GIS~\citep{grassgis}, both open-source, offer five methods each, whereas Mapbox Studio~\citep{mapbox} offers four.
Compared to these, \app{} offers 16+ binning methods.

From a purely styling standpoint, ColorBrewer~\citep{harrower2003colorbrewer}, web-based color wheels, and in-house color palettes are used to determine appropriate web-, print-, and colorblind-friendly color schemes. Web-based or contrast-checkers are used to ensure a color contrast between elements, such as geographic entities and their labels, meeting accessibility standards and enhancing readability for all users. Lastly, Illustrator~\citep{illustrator}, Photoshop~\citep{photoshop}, and Inkscape~\citep{inkscape} are used for blending, after effects, and other postprocessing tasks. \app{} provides many pre-defined color palettes and also lets users filter based on their web-, print, and colorblind-friendliness.

\paragraphHeadingSpace\bpstart{Libraries.} GIS libraries provide APIs to help developers build custom geospatial applications. Particularly for choropleth mapmaking, ArcGIS Maps SDK (Software Development Kit)~\citep{arcgisjssdk}--available in JavaScript, Java, Kotlin, .NET, QT, Swift, and Python programming languages--supports six binning methods, Python's PySAL~\citep{pysal} supports ten binning methods, and tmap~\citep{tmap} in R supports nine methods. BinGuru~\citep{narechania2023resiliency} is a recent, novel Javascript library that provides implementations of 18+ binning methods, using which we built \app{}.

\subsection{Exploratory Search \& Analysis and Comparison in Visualizations}
\bpstart{Exploratory Search \& Analysis.} 
Prior work on exploratory search~\citep{marchionini2006exploratory, white2009exploratory} and data analysis~\citep{heer2012interactive, tukey1977exploratory} involves two types of tasks: browsing and searching. Browsing helps gain an overview and engage in serendipitous discovery and search helps find answers to specific questions; this helps users who may be unfamiliar with the resources available (e.g., their data), are in the midst of forming goals, or are unsure of best means to achieve their goals.
A popular exploratory search approach is Faceted Browsing~\citep{yee2003faceted} to explore collections in which users specify filters using metadata to find subsets of items sharing desired properties.
Voyager~\citep{wongsuphasawat2015voyager} utilizes this technique to recommend charts based on statistical and perceptual measures. 
Voyager2~\citep{wongsuphasawat2017voyager} similarly blends manual and automated chart specification to help analysts engage in both open-ended exploration and targeted question answering.
We designed \app with a similar approach to browse and filter multiple binning methods.

\paragraphHeadingSpace\bpstart{Comparison.} To support designing for effective visual comparisons across binning methods, we used principles from literature on visual comparison~\citep{gleicher2011visual, gleicher2017considerations}, set visualization~\citep{alsallakh2016state}, composite visualization views~\citep{javed2012exploring}, and coordinated multiple views~\citep{roberts2007state}. 
This visual comparison taxonomy comprises three categories: \emph{juxtaposition}, \emph{superposition}, and \emph{explicit encoding of relationships}; the composite visualization design space introduces five patterns: \emph{juxtaposition}, \emph{superimposition}, \emph{overloading}, \emph{nesting}, and \emph{integration}. 
Euler and Venn diagrams represent each set as a geometric shape and show the intersections by overlapping the shape, and remain the most common methods to visualize sets and their intersections. 
Onset~\citep{sadana2014onset} uses direct manipulation interaction and visual highlighting to support easy identification of commonalities and differences as well as membership patterns sets. UpSet~\citep{lex2014upset} visualizes set intersections in a matrix layout introducing aggregates based on groupings and queries. We designed \app to predominantly use the \emph{juxtaposition} design pattern and place the binning method outputs side-by-side (or stacked)~as small multiples.

\subsection{Choropleth Map Data Binning}
\bpstart{Number of Bins.}
Choropleth map binning enables pattern exploration across spatial distributions. The color encoding on the map enforces visual grouping of regions based on assigned bin categories. The number of data bins used on maps typically ranges from 3--7. Mersey (\citeyear{mersey1990colour}) recommends 5--9 bins, while Declerq (\citeyear{declerq1995choropleth}) suggests 7--8 for accurate choropleth maps. Axis Maps advises 3--7 bins as a `safe' range~\citep{axismaps}, and desktop mapping software often defaults to 4--6 bins~\citep{declerq1995choropleth}. Brewer and Pickle's (\citeyear{brewer2002evaluation}) evaluation of binning methods was based on two experiments with 5 and 7 bins, respectively. Declerq's (\citeyear{declerq1995choropleth}) evaluation showed that 50\% of published choropleth maps use five bins, and 20\% use six bins. Olson (\citeyear{olson1972effects}) found that maps with more bins appear more similar regardless of method, and MacEachren (\citeyear{maceachren1982role}) noted that while detailed maps are interpretable, readers often reduce information to three ordinal categories (high, medium, low).

\paragraphHeadingSpace\bpstart{Binning/Classification Methods.}
A cornerstone of this work is displaying a variety of binning methods. After searching literature and software documentation, we sourced a set of binning methods and divided them into six high-level categories based on how they compute or determine bins: \emph{interval-based}, \emph{statistical}, \emph{iterative}, \emph{spatial}, \emph{human-centered}, and \emph{others}. We provide a subset of these methods in \app{}, and outline them in Table~\ref{tab:SummaryMethods}. Additional descriptions, references, examples, and usage guideline(s) associated with these binning categories and methods are available as part of Supplemental Material.

% \bigskip
% \textbf{---Table 1 near here ---}
% \bigskip
\begin{table}
    \centering
    \small
    \setlength{\tabcolsep}{4pt}
    \renewcommand{\arraystretch}{1.5}
    \caption{Overview of \textbf{Categories} of Binning Methods for Choropleth Mapmaking including \textbf{Definitions}, and \textbf{Examples}, highlighting the ones \emph{currently available in \app{}}.}
    \label{tab:SummaryMethods}
    \begin{tabular}{@{} p{1.25cm} p{6cm} p{6cm} @{}}

        \toprule

        \textbf{Category} & \textbf{Definition} & \textbf{Examples} \emph{available in \app{}} \\
        
        \midrule

        Interval-based & Methods that create arithmetically-generated breaks to divide sets of values. & \emph{Equal Interval}, \emph{Defined Interval}, \emph{Geometric Interval}, \emph{Exponential}, Logarithmic, \emph{Maximum Breaks}\\
        
        Statistical & Methods that use summarized descriptive values about the data to create groups. & \emph{Quantile}, \emph{Percentile}, \emph{Box Plot}, \emph{Mean - Standard Deviation}, Nested Means \\
        
        Iterative & Methods that follow a process or algorithm to reach an objective function & Clustering-based Algorithms (K-Means, \emph{CK-Means}, DBSCAN), \emph{Head-Tail Breaks}, \emph{Natural Breaks}, Max-P \\
        
        Spatial & Methods that incorporate ancillary geographic information or geometric constraints. & Equal Area Breaks / Geometric Quantile, Shared Area, Minimum Boundary Error, Overview Accuracy Index (OAI), Boundary Accuracy Index (BAI) \\
        
        Human-Centered & Methods that focus on legibility and interpretability. & \emph{Manual Interval}, \emph{Pretty Breaks}, Semantic Binning \\
        
        Other & Other important methods that do not fall into the above categories. & \emph{Unclassed (Continuous)}, Unique Values, \emph{Resiliency}, Equal Interval Size Area, Equiprobabilities, Discont., Hybrid Equal Interval-Box \\

        \bottomrule
    \end{tabular}
\end{table}

\paragraphHeadingSpace 
In this section, we reviewed key background knowledge used to inform the design of a tool called Exploropleth that helps users compare and choose binning methods for choropleth maps. This tool integrates the information shared in this section in several ways. First, it uses a default number of bins/classes (n=5) that aligns with prior suggestions. Next, it uses a default color ramp that helps uses differentiate between bins/classes. Following, the types of binning methods are populated from the literature review and are displayed in ascending order based on the top-down grouping by type of technique and methods more commonly found in the literature. We also integrate this set of literature into other `tabs' on the tool. Because histograms are also a common way of viewing data, we include a tab that focuses solely on histogram comparison. Because manual classification is a popular and important way of classifying data, we include a tab that allows users to begin with a canonical binning method (specifically, natural breaks), and then tweak the bins from this starting point. 
In the next section, we outline the design goals of \app{} in more detail.

\section{\app}
\label{section:userinterface}
\subsection{Key Terminologies}
During our review, we encountered certain terminology referring to the same concept, so for consistency, we first define these key terminologies and then describe \app.

\paragraphHeadingSpace\bpstart{Administrative units} refer to the geographic areas or regions on a map representing how a land may be divided, e.g., state, county, city.

\bpstart{Bin} refers to the single range of continuous values used to group values, also known as `class', `scheme', `category', `bucket', `break', `range'.

\bpstart{Bin count} is the number of specified bins for or created after binning.

\bpstart{Bin extent} refers to a bin's minimum and maximum values.

\bpstart{Bin interval} refers to the difference between a bin's maximum and minimum values. It is also referred to as `bin range'.

\bpstart{Bin size} refers to the number of data values that lie within a bin extent.

\bpstart{Bin breaks} refers to the set union of all bin extents.

\bpstart{Bin color} is the map color assigned to administrative units in that bin.

\subsection{Design Goals}
Considering the diversity and complexity of established binning methods documented in cartographic literature, along with the capabilities and challenges of existing geospatial visualization tools and libraries, we derived six goals to guide the design and development of \app.

\paragraphHeadingSpace\bpstart{G1. Showcase multiple binning methods.} 
We derived this goal to enable users to view and analyze multiple binning methods in the same place at the same time.

\paragraphHeadingSpace\bpstart{G2. Facilitate comparisons between multiple binning methods.} We derived this goal to let users compare binning methods based on their resultant bin counts, sizes, and intervals.

\paragraphHeadingSpace\bpstart{G3. Analyze combinations of established binning methods.} We derived this goal to let users analyze how administrative units get placed in different bins based on the binning method.

\paragraphHeadingSpace\bpstart{G4. Support manual classification and on-the-fly administrative unit reclassification.} We derived this goal to enable users to customize bins by specifying bin counts and bin intervals and/or reclassifying them by forcing administrative units into specific bins.

\paragraphHeadingSpace\bpstart{G5. Afford integrations into existing mapmaking workflows.} We derived this goal to enable mapmakers to integrate \app into their existing workflows with affordances to import their own data and also export resultant map artifacts.

\paragraphHeadingSpace\bpstart{G6. Enable user control and context through configurability.} We derived this goal to enable users to customize established binning methods (e.g., tweak input parameters such as bin count, bin interval) and the resultant maps (e.g., styling affordances such as color schemes).

\paragraphHeadingSpace\bpstart{G7. Support data exploration.} We derived this goal to enable users to independently analyze the (up)loaded dataset before analyzing and interacting with the data binning methods.

\subsection{User Interface}
We designed the user interface (UI) with four tabbed views each navigable from others to \textbf{\texticon{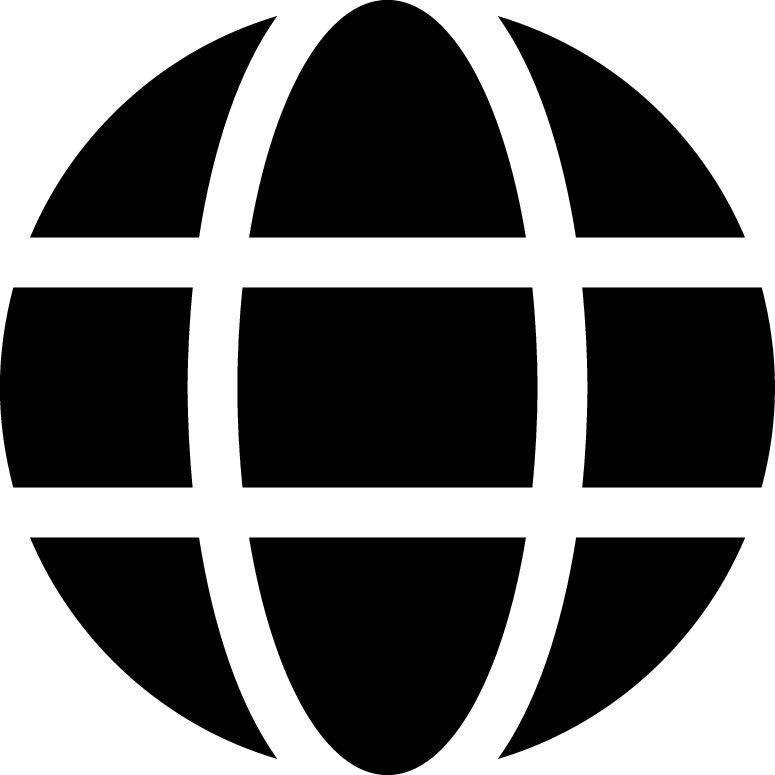}~Browse}, \textbf{\texticon{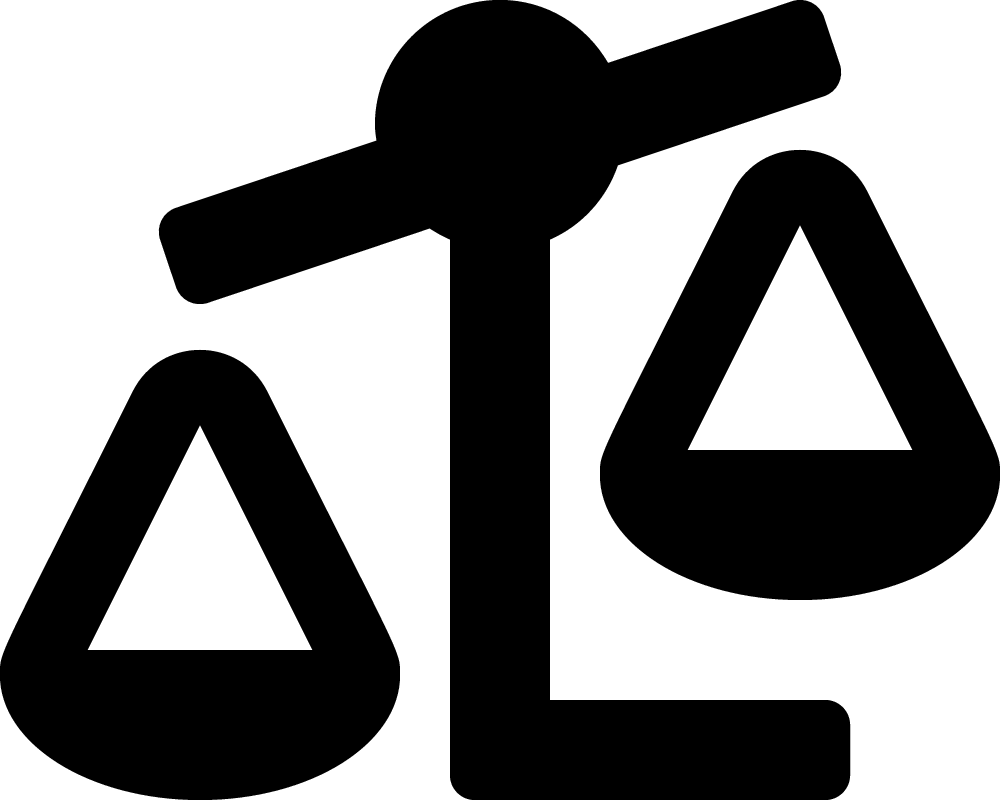}~Compare}, \textbf{\texticon{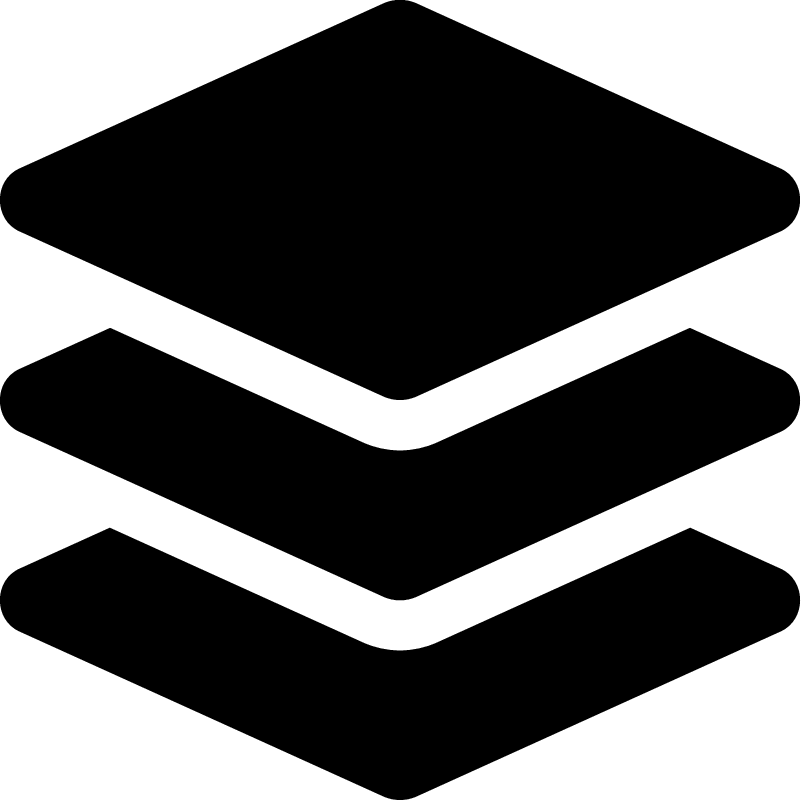}~Combine}, and \textbf{\texticon{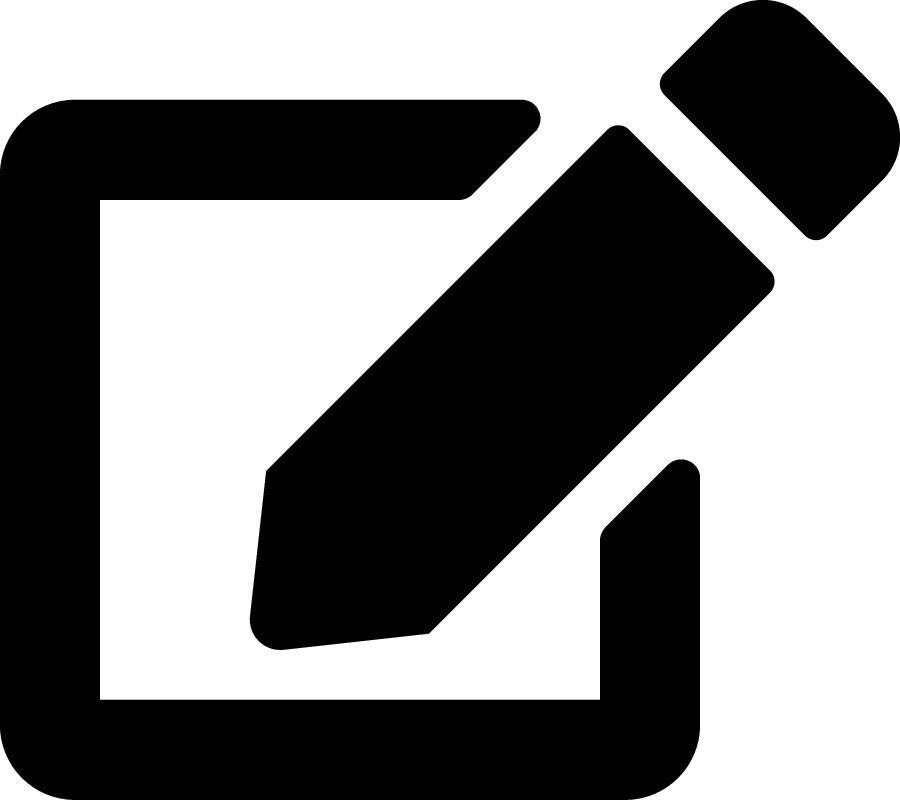}~Create} binning methods and a shared \textbf{Data and Configurations View}.
We finalized this design after discussing alternate designs with two visualization experts as pilot users (Section~\ref{subsection:pilotstudies}) and refining it based on feedback from sixteen GIS experts (Section~\ref{section:expert-interviews}).

% \bigskip
% \textbf{---Figure 1 near here ---}
% \bigskip
\begin{figure*}[ht]
    \centering
    \includegraphics[width=\textwidth]{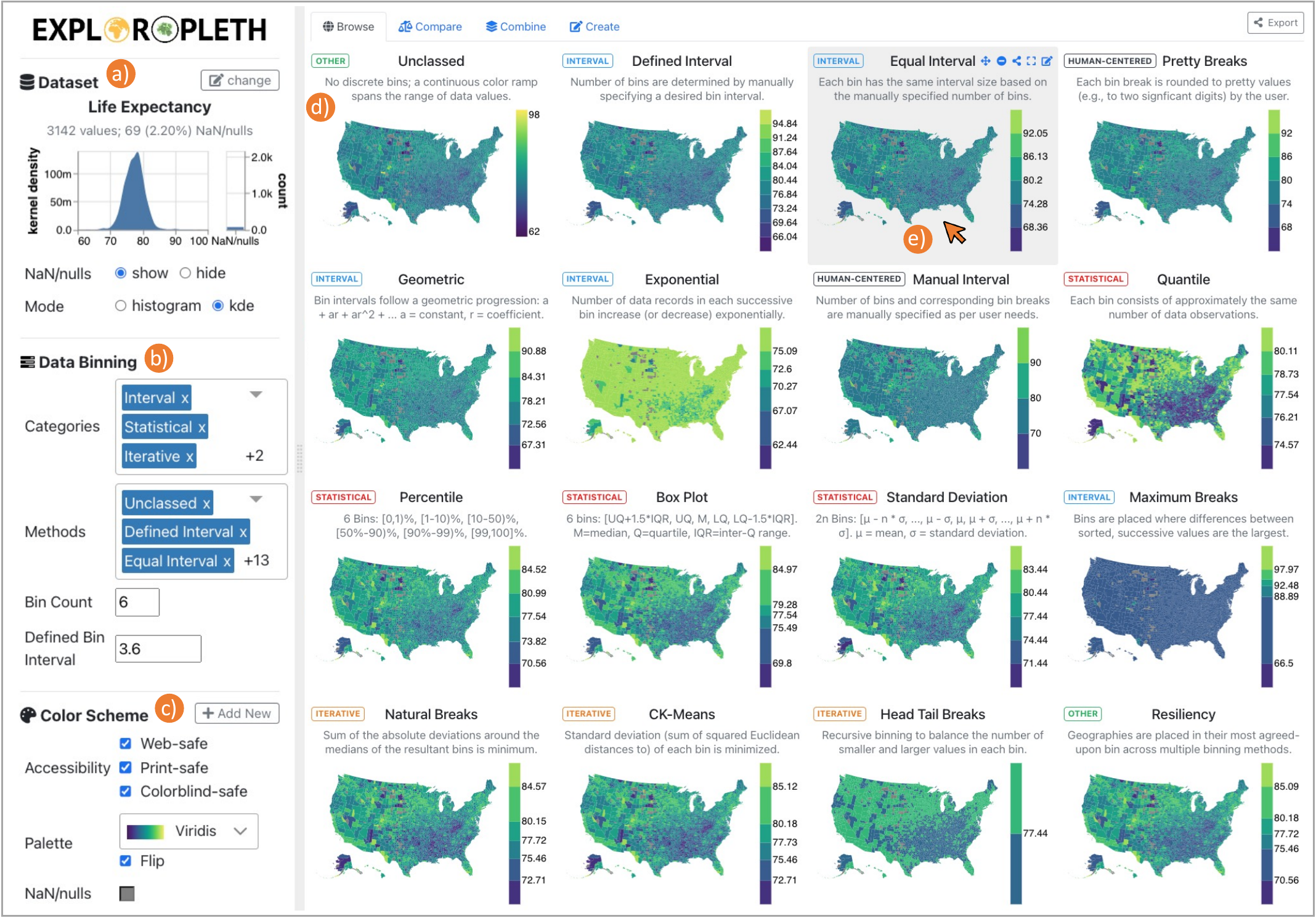}
    \caption{\emph{Browse View} showing a catalog of the same dataset and attribute across different data binning strategies. Users can a) configure the dataset, b) select binning methods and categories to explore, c) chose a color scheme, d) visualize the resulting binning outputs, and e) inspect details by interacting with a specific output.}
    \label{fig:browse}
\end{figure*}

\subsubsection{Data and Configurations View}
This view enables users to load their data, analyze the data distribution, and configure binning method parameters and map styles~(\textbf{G6}); it consists of three subviews~(Figure~\ref{fig:browse}):

\paragraphHeadingSpace\bpstart{\texticon{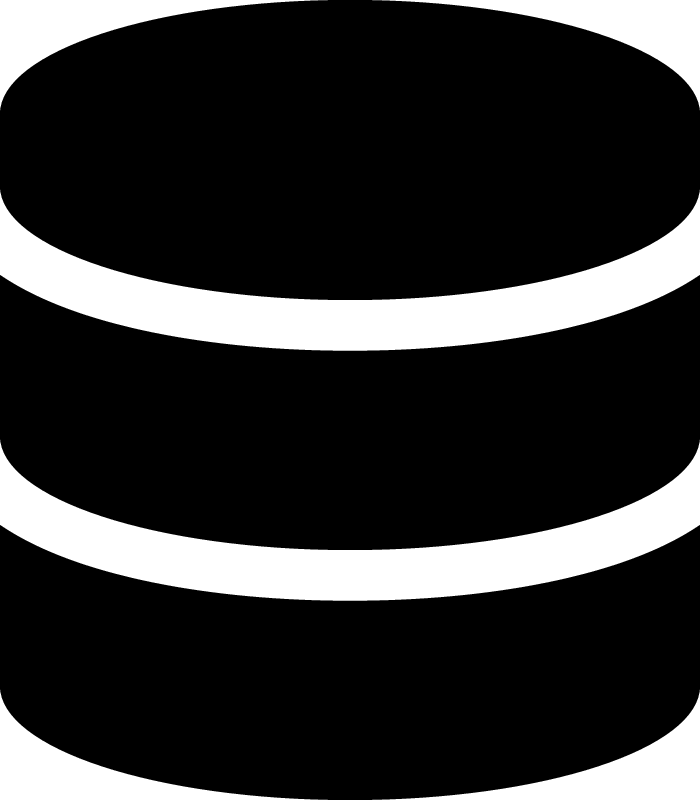}~Dataset.} This subview (Figure~\ref{fig:browse}a) lets users~\texticon{edit-solid}~\emph{change} the selected geography (e.g., United States counties) or feature dataset (e.g., \emph{Life Expectancy}) by either uploading their own files or by choosing from available samples.
For the loaded dataset feature, a summary data profile is shown along with a visualization of the underlying data distribution (histogram or kernel density estimate plot) with options to toggle missing (NaN/null) values~(\textbf{G7}).

\paragraphHeadingSpace\bpstart{\texticon{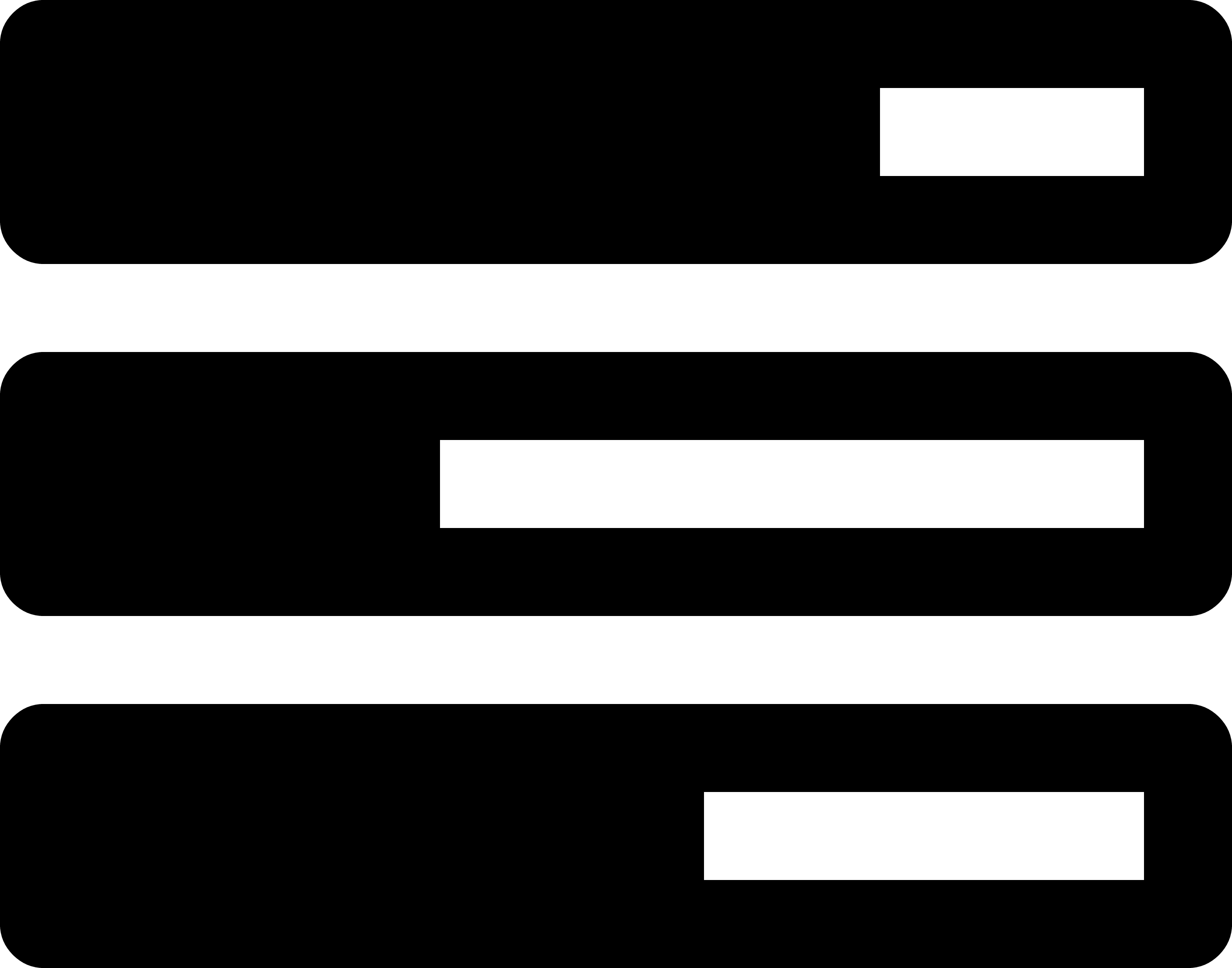}~Data~Binning.} This subview (Figure~\ref{fig:browse}b) provides options to filter the adjacent views (\emph{Browse}, \emph{Compare}, \emph{Combine}) based on sixteen currently supported data binning methods (\emph{unclassed, defined interval, equal interval, pretty breaks, geometric interval, exponential interval, manual interval, quantile, percentile, box plot, standard deviation, maximum breaks, natural breaks, ck-means, head-tail breaks, and resiliency}) and their five corresponding binning categories (\emph{human-centered}, \emph{interval-based}, \emph{statistical}, \emph{iterative}, \emph{other}).
We currently support only these methods (and categories) due to their popularity and algorithmic availability and feasibility.
The ``Bin Count'' and ``Defined Interval Size'' parameters can modify the \emph{dependent} binning methods that rely on them as inputs (e.g., modifying ``Bin Count'' will update \emph{equal interval} but not \emph{defined interval}, which can be modified by ``Defined Interval Size'').

\paragraphHeadingSpace\bpstart{\texticon{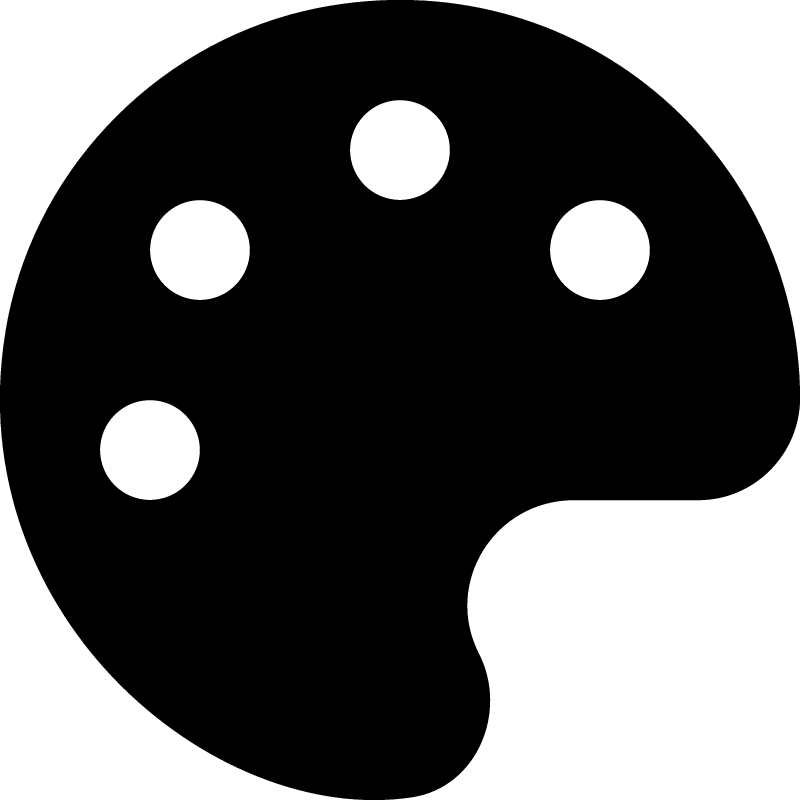}~Color~Scheme.} This subview (Figure~\ref{fig:browse}c) provides options to choose an appropriate color scheme for valid as well as missing (NaN/null) values across the maps in the adjacent views. Numerous color palettes across five types of color scales are available: \emph{categorical}, \emph{sequential single-hue}, \emph{sequential multi-hue}, \emph{diverging}, and \emph{cyclical} along with an option to use them in \emph{reverse}. Users can filter these color scales and palettes to make accessible maps (\emph{\{web, colorblind, print\}}-friendly). Users can also \texticon{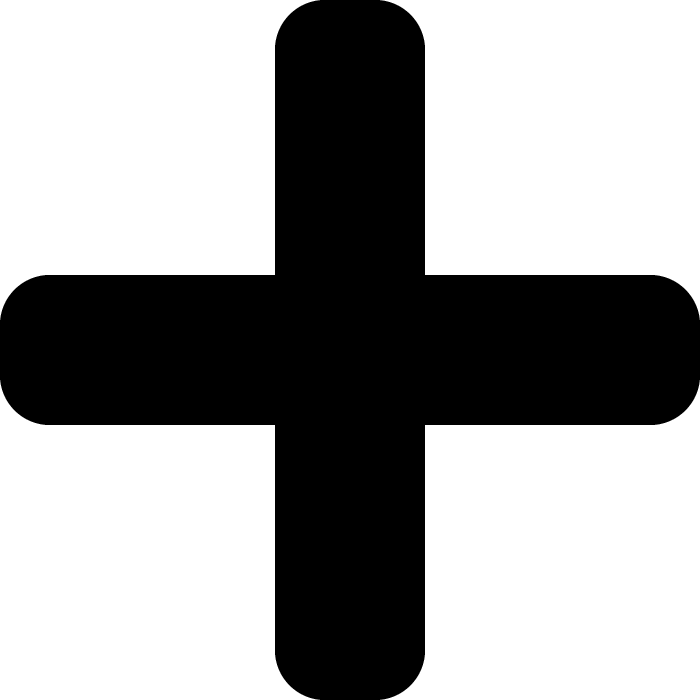}~\emph{add~new}, custom color palettes.

\subsubsection{Browse View}
% \bigskip
% \textbf{---Figure 2 near here ---}
% \bigskip
\begin{figure}[ht]
    \centering
    \includegraphics[width=0.6\columnwidth]{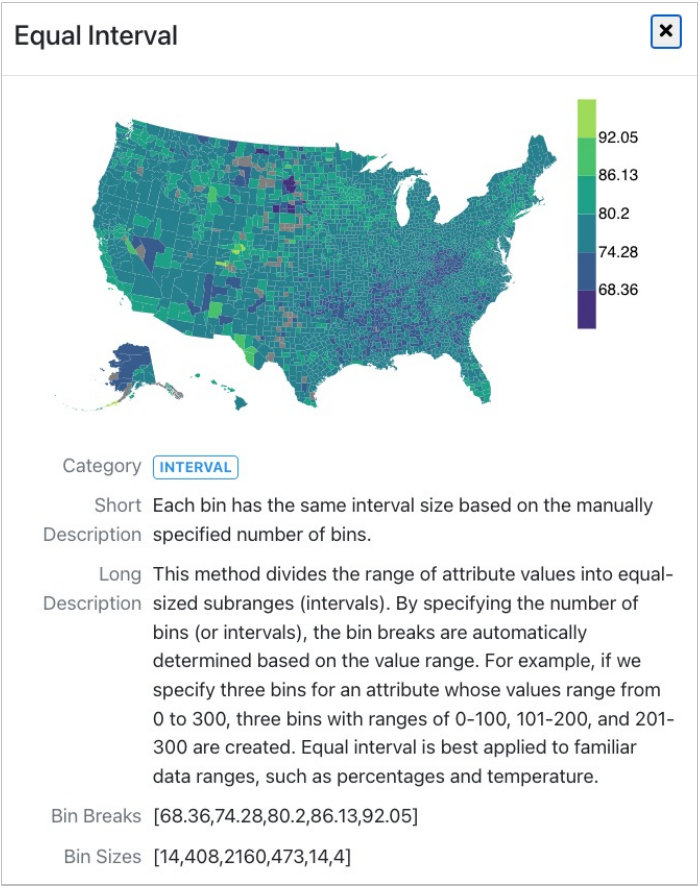}
    \caption{\emph{Detail View} showing a choropleth with descriptions of the binning method used and resultant bin breaks and sizes.}
    \label{fig:detailview}
\end{figure}

\begin{figure}[ht]
    \centering
    \includegraphics[width=0.6\columnwidth]{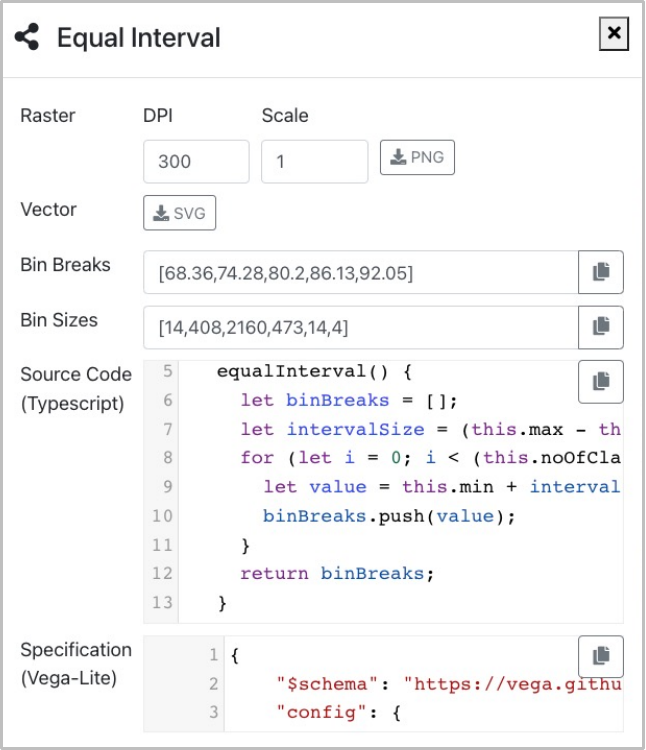}
    \caption{\emph{Export View} offers affordances to help users export the map in vector as well as raster formats, copy the resultant bin breaks, bin sizes, corresponding source code and the visualization specification.}
    \label{fig:exportview}
\end{figure}

This view~(Figure~\ref{fig:browse}d) shows small multiples~\citep{van2013small} of sixteen cards in a grid layout, each with a choropleth map showing the same administrative units and data but colored based on different binning methods~(\textbf{G1}). Each card (Figure~\ref{fig:browse}e) shows the corresponding binning method's name, a short two-line description and interactive affordances to: \textbf{\texticon{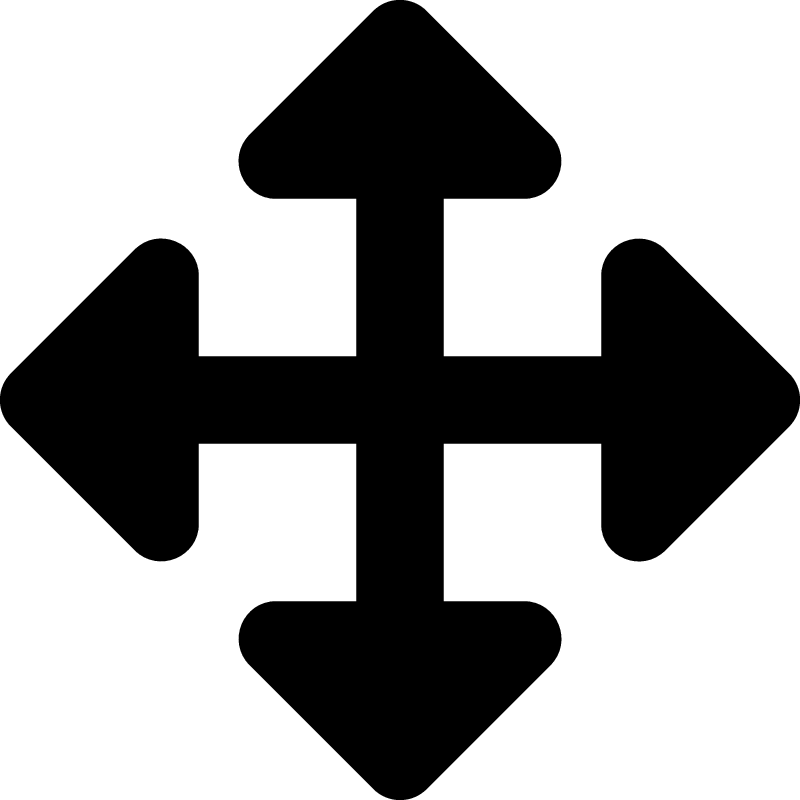}~drag} and reorder cards to facilitate comparisons~(\textbf{G2}), temporarily \textbf{\texticon{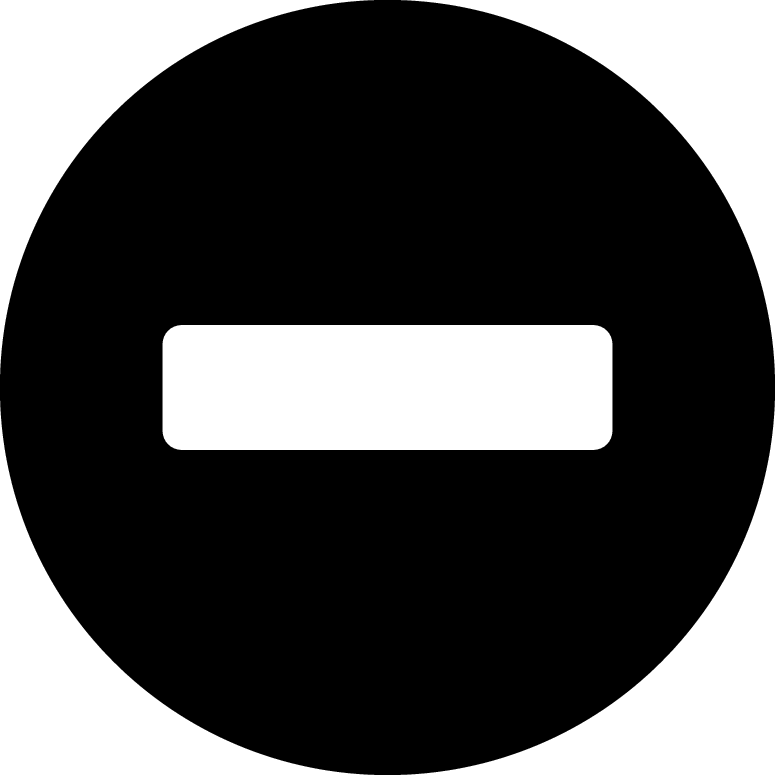}~hide} undesirable binning methods~(\textbf{G6}), \textbf{\texticon{edit-solid}~customize} the output of the binning method in the \emph{Create View}~(\textbf{G4}), \textbf{\texticon{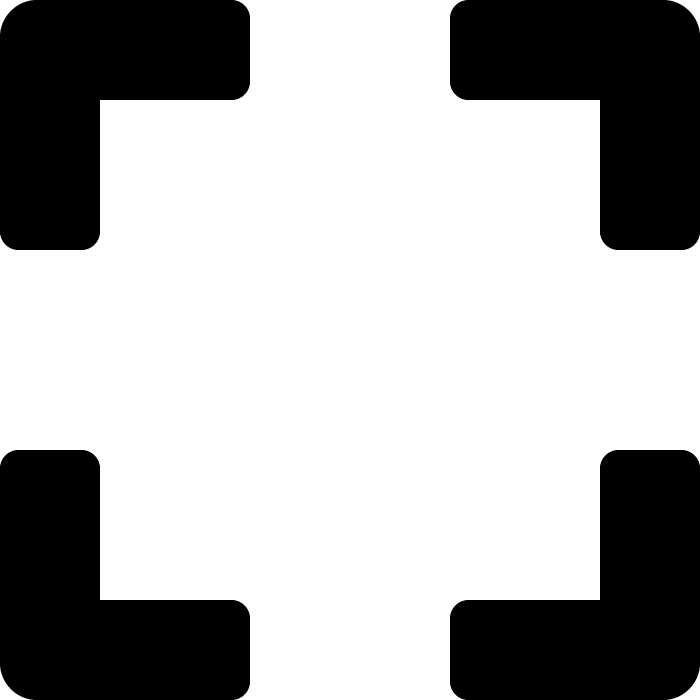}~expand} into a larger, detailed view (Figure~\ref{fig:detailview}), and \textbf{\texticon{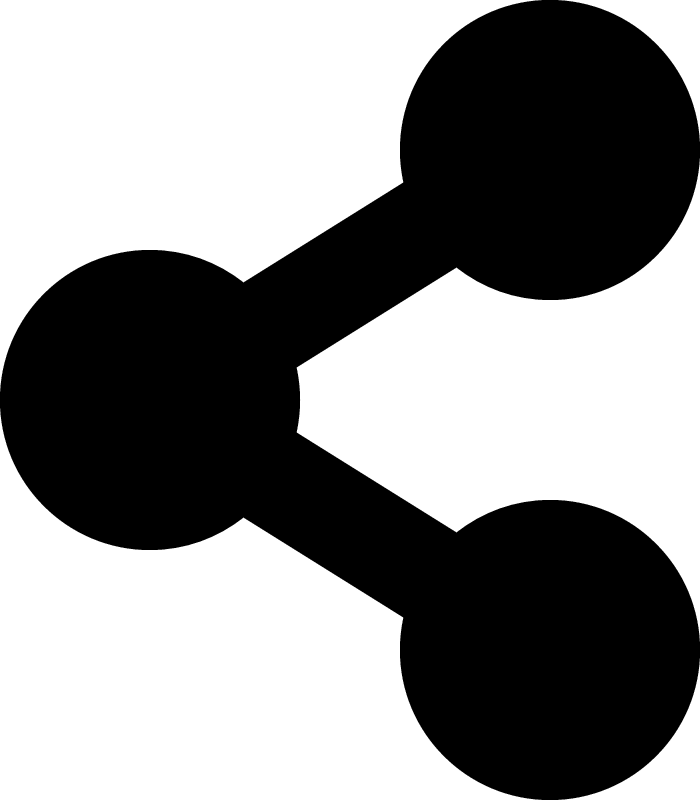}~export} the card as an PNG/SVG image across different scales and resolutions, the binning method's source code in Typescript, and the choropleth map's visualization specification in Vega-Lite (Figure~\ref{fig:exportview},~\textbf{G5}).

% \bigskip
% \textbf{---Figure 3 near here ---}
% \bigskip

With these affordances, users can visualize the variance in binning method outputs based on the geographic distribution of colors. For example, the choropleth corresponding to \emph{quantile} appears to have an equal distribution of blue and green shaded counties, as by definition, the method places an approximately equal number of data points in each bin, whereas the choropleth corresponding to \emph{maximum breaks} appears mostly blue~(Figure~\ref{fig:browse}).

\subsubsection{Compare View}

% \bigskip
% \textbf{---Figure 4 near here ---}
% \bigskip
\begin{figure}[!h]
    \centering
    \includegraphics[width=0.55\columnwidth]{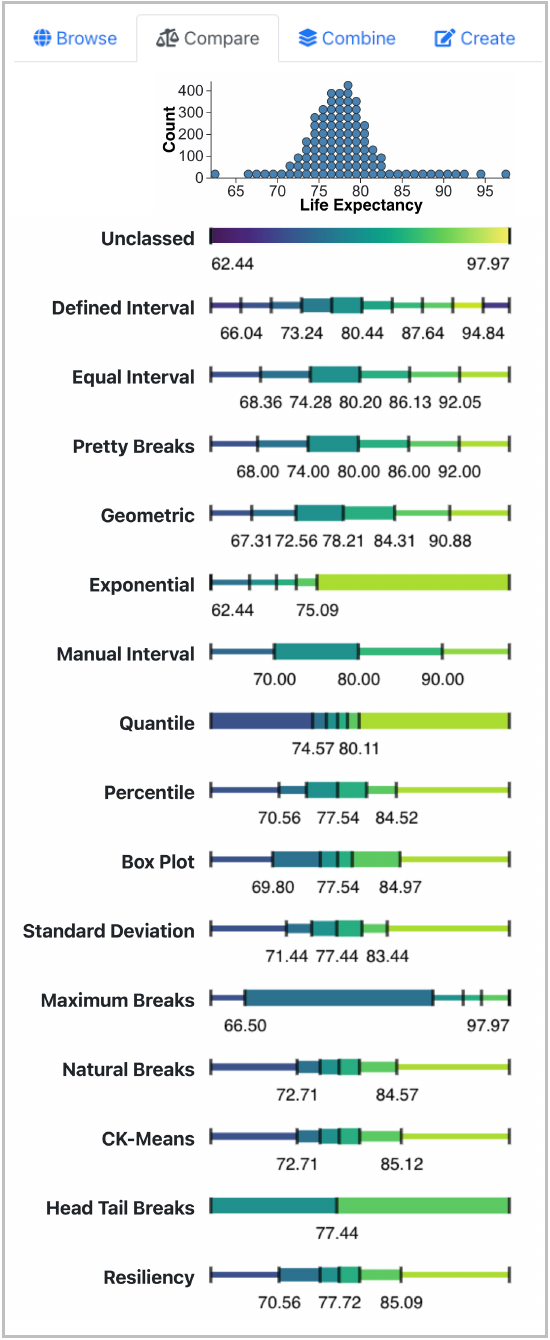}
    \caption{\emph{Compare View} shows the underlying data distribution (as a dot plot) and small multiples of established and custom binning method outputs (as bar charts), the latter facilitating a visual comparison based on the resultant bin counts (black ticks), intervals (bar length), number of items in each bin (bar height).}
    \label{fig:compare}
\end{figure}

This view~(Figure~\ref{fig:compare}) visualizes the underlying distribution of the attribute values as an interactive dot plot, and the binning methods as small multiples stacked on top of each other to enable comparisons based on the resultant (or specified) bin count, intervals, and sizes~(\textbf{G2}).
Users can interactively drag and reorder these small multiple visualizations to promote effective comparisons.
Each small multiple visualization corresponds to a binning method and is designed as a horizontal stacked bar chart with the bin count, interval, and size encoded as the stack count, width, and height, respectively.
Bins are colored based on the chosen color scheme and bin breaks are made prominent with black, vertical strips.
If a bin has no data values, then a dashed black line connects the adjacent bins, conveying zero features in that bin.
Hovering on top of a bin displays a tooltip showing its corresponding size, extent, and interval.

With these affordances, users can visualize the variance across binning method inputs and outputs in terms of the bin counts, sizes, and intervals. For example, all stacks in the \emph{quantile} method are of the same height (but varying widths), as by definition, the method ensures an approximately equal number of data points in each bin (Figure~\ref{fig:compare}).

\subsubsection{Combine View}
\label{subsection:combine}

% \bigskip
% \textbf{---Figure 5 near here ---}
% \bigskip
\begin{figure*}[ht]
    \centering
    \includegraphics[width=\textwidth]{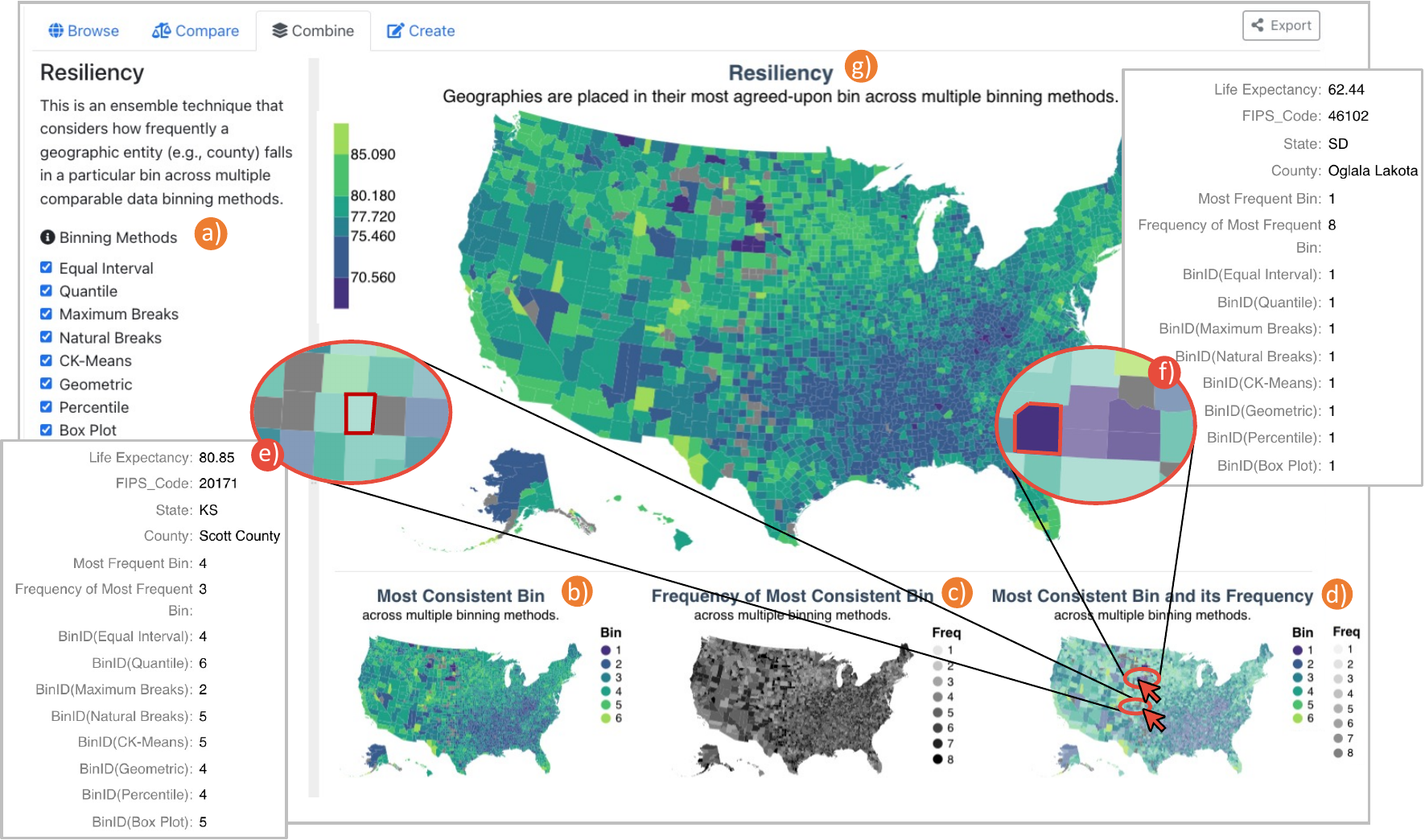}
    \caption{\emph{Combine View} lets users analyze a) a combination of one or more binning methods by visualizing b) the \emph{Most Consistent Bin}, c) the \emph{Frequency of the Most Consistent Bin}, and d) both together for each U.S. county in separate choropleths. The new, \emph{Resiliency} binning method then utilizes this information to g) determine ``resilient'' bins (counts, intervals) that are also visualized in a choropleth. Hovering any county on the map shows a tooltip with relevant information about the county, as shown in e) and f).}
    \label{fig:combine}
\end{figure*}

This view~(Figure~\ref{fig:combine}) enables users to combine two or more binning methods and visualize combination representations, including interacting with the \emph{resiliency}~\citep{narechania2023resiliency} ensemble binning method~(\textbf{G3}).
For each administrative unit, we first track the index of the bin (\emph{binID}) it was placed in across eight binning methods (Figure~\ref{fig:combine}a): \emph{equal interval}, \emph{quantile}, \emph{maximum breaks}, \emph{natural breaks}, \emph{ck-means}, \emph{geometric interval}, \emph{box-plot}, and \emph{percentile}. 
We chose these methods as all methods have six bins, either by definition (\emph{box-plot}, \emph{percentile}) or through the ``Bin Count'' configuration in the \emph{Data and Configuration View} (the other six methods). 
It is desirable to have the same bin count to combine binning methods.
Next, we compute the frequency of these \emph{binID}s. 
For example, Scott County, Kansas has the following \emph{binID}s across the eight methods: \emph{equal interval} (4), \emph{quantile} (6), \emph{maximum breaks} (2), \emph{natural breaks} (5), \emph{ck means} (5), \emph{geometric interval} (4), \emph{percentile} (4), and \emph{box plot} (5); the most frequent bin (4); and the frequency of the most frequent bin (3), as shown in~Figure~\ref{fig:combine}e. We present these intermediate results visually through three choropleth maps:

\paragraphHeadingSpace\bpstart{Most Consistent Bin.} This map shows the \emph{most frequent (consistent) bin} across binning methods for each administrative unit and is colored using a qualitative color scheme~(Figure~\ref{fig:combine}b).

\paragraphHeadingSpace\bpstart{Frequency of Most Consistent Bin.} This map shows the \emph{frequency of its most frequent (or consistent) bin} across binning methods for each administrative unit and is colored using a sequential color scheme of blacks with higher frequencies as darker shades~(Figure~\ref{fig:combine}c).

\paragraphHeadingSpace\bpstart{Most Consistent Bin and its Frequency.} This map combines the previous two maps into a value-by-alpha map~\citep{roth2010value}. The color hue corresponds to the \emph{most frequent bin} and the opacity corresponds to the \emph{frequency of the most frequent bin}~(Figure~\ref{fig:combine}d), where higher opacity implies higher frequency For example, Oglala Lakota County, South Dakota (life expectancy = 62.44 years; lowest in the U.S.) in an opaque purple color is one of the most consistent counties in $\emph{binID}=1$~(Figure~\ref{fig:combine}f); Scott County, Kansas (life expectancy = 80.85 years) in a translucent green shade is one of the most inconsistent counties with $\emph{binID}s=\{2, 4, 5, 6\}$ across eight binning methods~(Figure~\ref{fig:combine}e). 

\paragraphHeadingSpace\bpstart{Resiliency.} These intermediate results are processed by the recent consensus binning method, Resiliency~\citep{narechania2023resiliency}, to determine new bin counts, extents, and sizes, and are visualized in another choropleth~(Figure~\ref{fig:combine}g). 

\subsubsection{Create View}

% \bigskip
% \textbf{---Figure 6 near here ---}
% \bigskip
\begin{figure*}[ht]
    \centering
    \includegraphics[width=\textwidth]{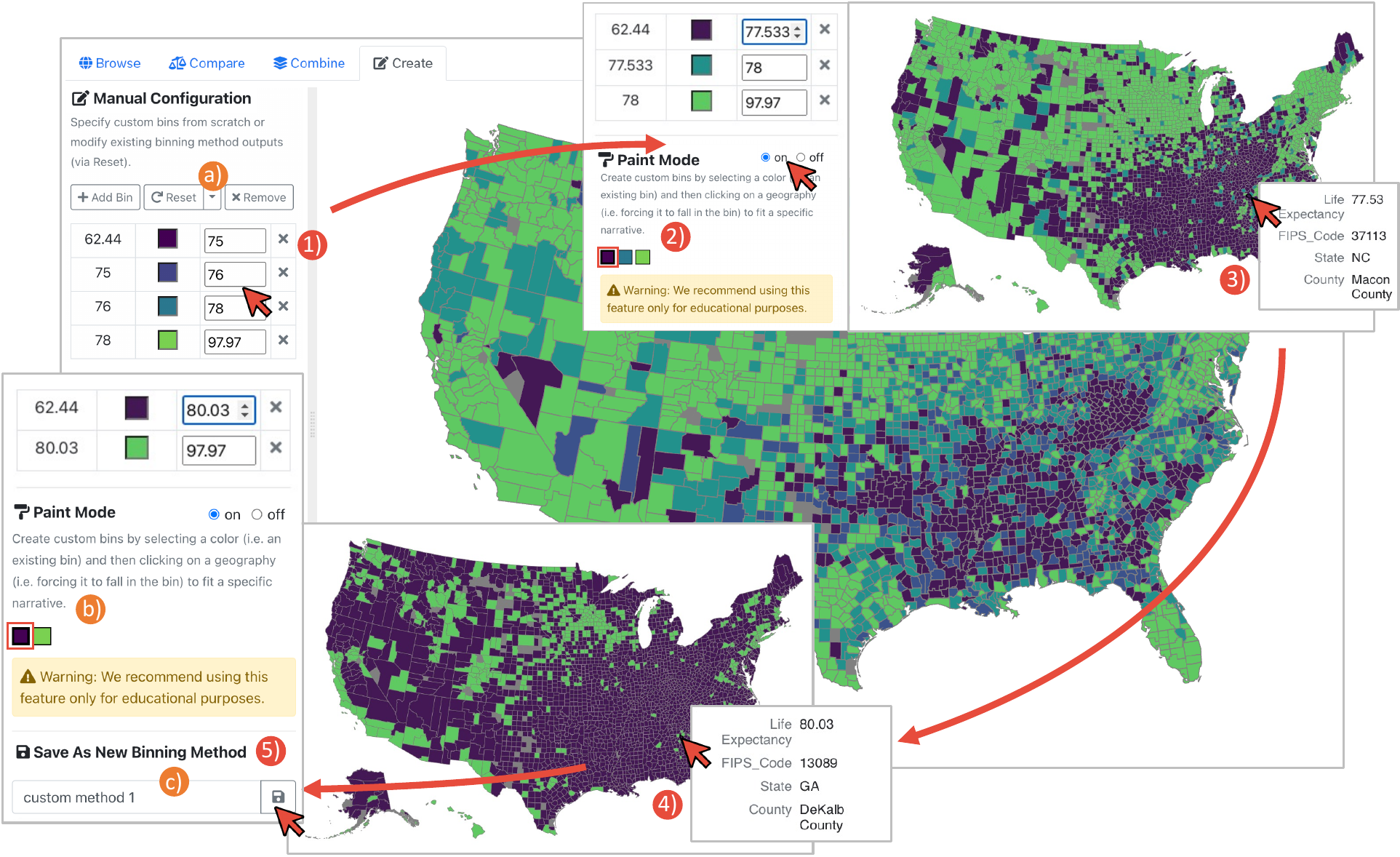}
    \caption{\emph{Create View} lets users create custom bins either from scratch, by a) modifying the resultant bins from one of the established binning methods or b) reclassifying administrative units into specific bins and c) save them for future use. 1)--5) illustrates how users can use 1) the manual binning specification controls and 2) the \emph{Paint Mode} to interactively make 3) Macon County and 4) Dekalb County, originally in different bins, to fall in the same bin, and 5) save them for later use.}
    \label{fig:create}
\end{figure*}

This two-columned view (Figure~\ref{fig:create}) contains a configuration panel for manually specifying bins and a canvas for visualizing the corresponding results in an interactive choropleth map in real time~(\textbf{G4}). The configuration panel comprises three subviews:

\paragraphHeadingSpace\bpstart{\texticon{edit-solid}~Manual~Configuration.} This subview~(Figure~\ref{fig:create}a) enables users to utilize direct manipulation~\citep{hutchins1986direct} interaction paradigm to create custom bins--via \textbf{\texticon{plus-solid}~add} new bins, \textbf{\texticon{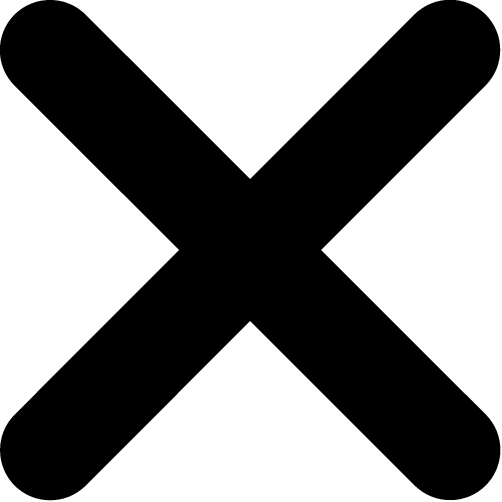}~remove} existing bins--or customize resultant bins of an existing binning method--via the \textbf{\texticon{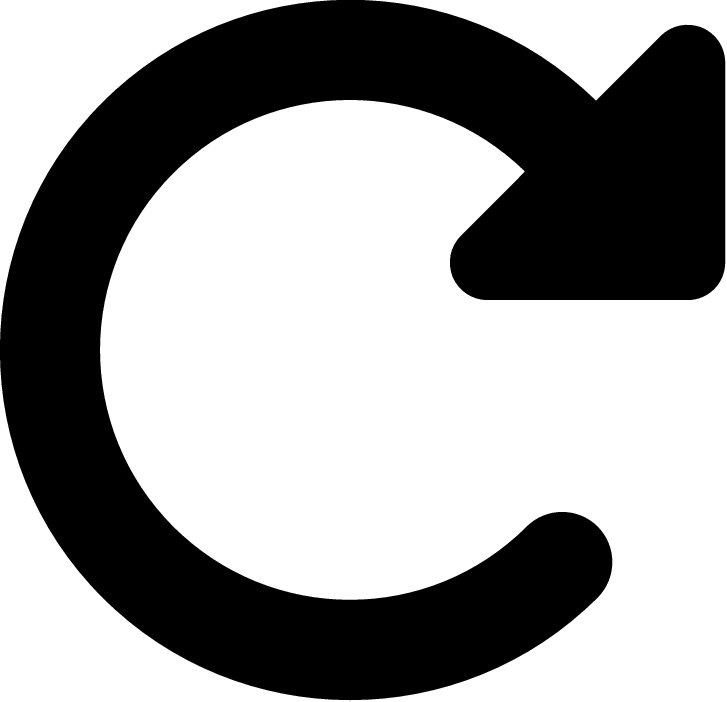}~reset} button and linked dropdown. Users can modify a bin's color and extent by interacting with the corresponding color swatch and numeric input fields, respectively.

\paragraphHeadingSpace\bpstart{\texticon{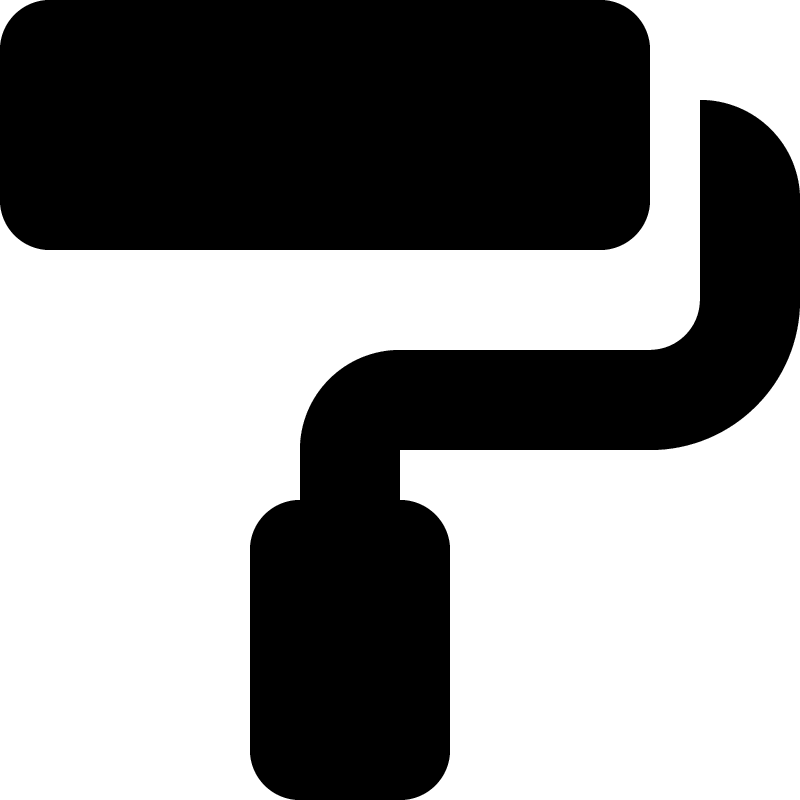}~Paint~Mode.} This subview~(Figure~\ref{fig:create}b) is a new way to reclassify administrative units by interactively assigning corresponding map administrative units to specific bins.
We call it this because this mode allows the mapmaker to ensure the map fits with their notion of where an entity should be binned to fit a narrative.
This reclassification can be performed on one of the established binning method outputs or also on the custom user-specified bins.
Users perform a \emph{visualization by demonstration} interaction~\citep{saket2016visualization} by first selecting the color corresponding to their target bin and then clicking on an administrative unit on the map to force it to fall into that bin. 
Our algorithm modifies the bin count and intervals to satisfy this bin assignment constraint. 
Users can specify multiple such constraints through multiple such `paint' operations.
However, to caution users against potential misuse or propaganda stemming from the creation of custom bins~\citep{monmonier2018lie}, this view also displays a warning: \emph{``We recommend using this feature only for educational purposes.''}

\paragraphHeadingSpace\bpstart{\texticon{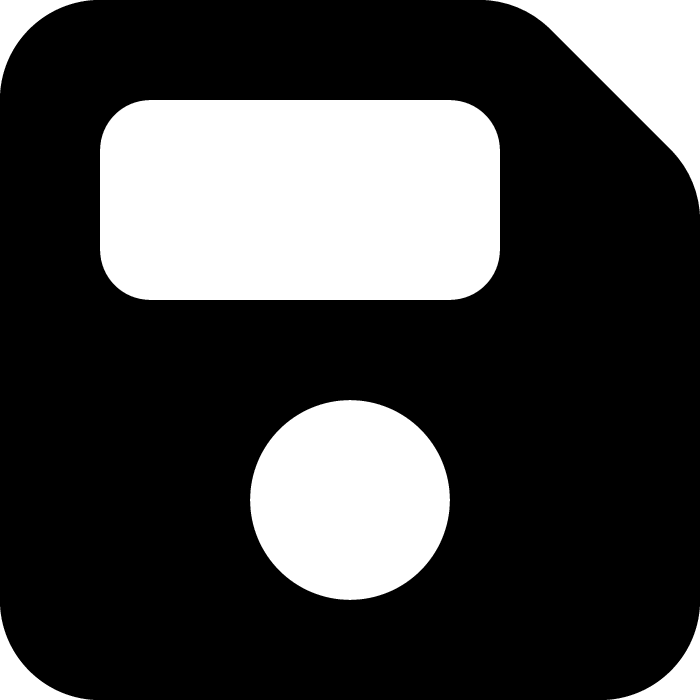}~Save~As~New~Binning~Method.} This subview~(Figure~\ref{fig:create}c) lets users \textbf{\texticon{floppy-disk-solid}~save} their customized set of bins; these bins are then also added to the \emph{Explore View} and \emph{Compare View} alongside visual representations of other methods to facilitate continued exploration and comparison.

\subsection{Implementation}
\label{subsection:implementation}
\app is built using the Angular~\citep{angular} framework. All current data binning methods are implemented using BinGuru~\citep{narechania2023resiliency},
visualizations are built using Vega-Lite~\citep{satyanarayan2016vega} or D3~\citep{d3}, and
color palettes are referenced from ColorBrewer~\citep{harrower2003colorbrewer}, d3-scale-chromatic~\citep{d3scalechromatic}, and Vega~\citep{vegacolorschemes}.

\subsection{Alternate Design Considerations and Pilot Studies}
\label{subsection:pilotstudies}
We showed an initial version of the tool to two pilot users (second-year PhD students studying data visualization). Based on their feedback, we modified existing features and also added new features and identified bugs. For example, we prototyped synchronous zooming and panning across all sixteen maps in the \emph{Browse View} but had to discard it as these interactions were very slow and not fluid~\citep{elmqvist2011fluid}, especially for dense, detailed geographies.

We also prototyped all sixteen cards to be part of one large canvas DOM element but discarded it as we could not support interactivity within and across maps through the \emph{drag-to-reorder}, \emph{hide}, \emph{export}, \emph{show details}, and \emph{edit} icon buttons; Vega-Lite currently does not support binding such input controls to concatenated views as part of the main canvas~\citep{satyanarayan2016vega}. 
Furthermore, the load time for one, large canvas was also very high.
Hence, our eventual implementation comprises 16 standalone DOM elements that are rendered on-the-fly.

We also considered visualizing the bin sizes in each \emph{Browse View} card as a separate legend to the left of each choropleth map or next to the existing legend, or as a combination legend that shows the bin extents and bin sizes as separate encodings.
However, our pilot users believed that both these approaches would clutter the user interface and due to their uncommonness, may confuse mapmakers and readers. Instead, we provide this information in the \emph{Compare View} by encoding the bin width and height to their interval and size, respectively.

\section{Usage Scenarios}
\label{section:examples}
% Scenario 1
\subsection{Scenario 1: Educating new mapmakers about binning methods.} Kiran is a student in an introductory GIS class (i.e., a domain novice and new to mapmaking) that uses \app for a class lab. For the lab, they open the app, select an existing dataset (e.g., \emph{``Life Expectancy''} in U.S. counties), inspect its data distribution, and start browsing the catalog of binning methods. For each method, they read its name and short description, and then visualize the resultant bin count, bin intervals, bin colors in the map legend (based on the viridis color scheme), and the distribution of the counties placed in each of the bins in the map.
To get details such as a longer description and precise bin breaks and bin sizes, they \textbf{\texticon{expand-solid}~expand} the card into a bigger view.

By visualizing these methods in one view, they discover that the \emph{quantile} method results in counties with an even distribution of blue and green shades (by definition) whereas the \emph{maximum breaks} method results in counties mostly colored in a dark shade of blue~(Figure~1, as there might be some outliers).
They also observe how certain methods have no bins (\emph{unclassed}), too few bins (\emph{head-tail breaks}), many bins (\emph{defined interval}), and a fixed number of bins (\emph{box plot}, \emph{percentile}) for this dataset. They modify the bin count and defined bin interval inputs from the \emph{Data and Configurations View} to see how certain binning methods update.
Hence, \app helped Kiran learn about different binning methods for choropleth maps.

% Scenario 2
\subsection{Scenario 2: Helping experienced mapmakers compare binning methods and choose one that suits their purpose.}
Vicky is an experienced mapmaker who makes choropleth maps in ArcGIS. Their current assignment is to prepare a choropleth map for the \emph{``Population Density''} of U.S. counties. They generally use the \emph{natural breaks} binning method, which is the ArcGIS default, or specify custom bins (\emph{manual interval}) for such tasks.
Because this dataset is heavily skewed--based on their inspection of the data profile in the \emph{``Data and Configuration View''}--\emph{natural breaks} was ineffective in highlighting certain densely populated urban counties such as ``New York County'' and ``Kings County''. 
They try the other six methods available in ArcGIS, and also specified custom bins but were not satisfied with the output.

They upload their dataset to \app and recognize the binning methods from ArcGIS, but find new methods such as \emph{head-tail breaks}. 
They see that this method effectively highlights densely populated outlier urban counties. 
They switch to the \emph{Compare View} to visualize bin details such as the sizes and extents, especially for the outlier counties.
After analyzing and comparing them, they select \emph{head-tail breaks} and \textbf{\texticon{share-alt-solid}~export} the corresponding \emph{bin breaks} (to manually create them in ArcGIS) and an \emph{image snapshot} to include in their task report.

% Scenario 3
\subsection{Scenario 3: Helping users reclassify administrative units on-the-fly.}
Shubham is a U.S. government official who wants to analyze \emph{life expectancy} statistics at a county-level to inform future policies.
They upload their data on \app and inspect the histogram of \emph{life expectancy} values.
Next, they switch to the \emph{Create View}, as they do not generally use established binning methods and often manually create custom bins.
They start customizing the pre-selected \emph{natural breaks} output (of six bins) by rounding off decimal points from bin extents and \textbf{\texticon{xmark-solid}~removing} a couple bins.
Next, they observe the map and notice many different shades of blue in the South (e.g., Georgia-GA, North Carolina-NC).  Per their domain knowledge, they assume some counties in the region have similar populations and low life expectancy and want them binned together in a ``lowest'' bin.
They enable \app's \emph{Paint Mode}, then they select (click on) the darkest shade of blue ($\emph{binID}=1$), and click on two counties that appear similar: ``Macon County'' (NC) and ``DeKalb County'' (GA).
In real time, \app modifies the bin counts and intervals, and updates the view. Most U.S. counties have now been grouped together to have the same dark blue color. Shubham also notices that the range is too large to show a meaningful public health pattern. 
They \textbf{\texticon{floppy-disk-solid}~save} their custom binning method as \emph{``South versus rest of the U.S.: Test 1''} and \textbf{\texticon{share-alt-solid}~export} the map to share with their colleagues as an example of the consequences of overly-broad data binning.
Figure~6~(1-5) illustrates this sequence of interactions in \app.

\section{Evaluation: Feedback from GIS Experts}
\label{section:expert-interviews}
To understand how \app can help mapmakers in their choropleth mapmaking process, we interviewed sixteen cartographers and GIS expert practitioners and researchers (\expert{1-16}) with a self-reported mapmaking experience between 3-33 years (median: 15 years, mean: 15.43 years). These experts worked with global government and non-government organizations like the World Health Organization (WHO) and federal agencies like the Department of Agriculture based in the U.S. (eleven) and India (two).
We primarily recruited these experts through their organizations' \emph{contact us} email addresses and form equivalents on their website.
Sessions were conducted remotely via Zoom and lasted between twenty five and thirty minutes.
There was no compensation for participation. This study was approved by the Georgia Tech Institute Review Board in the United States (Protocol Number: H22470).

Before each session, experts first provided verbal consent to participate in and be (screen, audio) recorded during the session. During the session, one administrator shared their screen and demonstrated \app for approximately ten minutes. A link where the tool was hosted was also shared with the experts to optionally interact during or after the demonstration; experts were encouraged to interrupt the administrator during the demonstration and think aloud while interacting by themselves. After the demonstration, the administrator facilitated a conversation for approximately fifteen minutes structured around the following questions:

\begin{enumerate}[nosep]
    \item \emph{Which data binning methods did you (not) know of?}
    \item \emph{Do you have anything specific to share about one or more of the binning methods that you learned about today?}
    \item \emph{(How) do you think a tool like this might be able to help you, other mapmakers, or other users?}
    \item \emph{What additional features would you like to see in this tool?}
\end{enumerate}

The session ended with experts filling out a brief background questionnaire. 
Due to restrictions at one expert's workplace, their interview was conducted over a phone call; we emailed relevant links and demonstrated the tool such that they were able to simultaneously interact and follow along the administrator's demonstration on their computer.

\subsection{Analysis}
\label{subsection:analysis}
We transcribed the audio recordings and segmented the transcripts into smaller units. We then conducted open coding~\citep{boyatzis1998transforming} wherein we applied `constant comparison'~\citep{strauss2008basics} by systematically comparing each segment with others to identify similarities and assign similar segments to common categories. We then grouped these categories into broader themes. We also employed `theoretical sampling'~\citep{strauss2008basics} to identify categories that had not yet emerged and refined them through iterative discussions within the research team. We followed this procedure throughout the evaluation process.

\subsection{General Feedback on \app}

\paragraphHeadingSpace\bpstart{\emph{```Nothing better than this' can happen for GIS experts \& coders.''}--\expert{3}}

\noindent All experts appreciated having multiple binning methods in one view, vindicating our initial assessment about the need for such a tool.
\expert{8} said, \emph{``It's amazing just how much you can change a map by changing their binning method.''}
\expert{2} found \app to be \emph{``a useful, lightweight tool that is intuitive and easy to maneuver.''}
\expert{14} suggested \app solves one of their challenges with existing tools: \emph{``If I want to change the bins, I have to go back and redefine [and replace] them. 
I cannot compare side by side. I have to export them [one at a time] which is just a whole hassle, and I can't imagine doing that for [sixteen] different types of binning methods. So having this is really cool.''}
Some experts noted their struggles with existing Python libraries \emph{``because [they are] not a computer programmer''} (\expert{16}) or \emph{``it is tedious to one-by-one explore the binning methods that are not supported by ArcGIS so having them in the same place is phenomenal''} (\expert{5}).
\expert{3} acknowledged \app would alleviate organizational challenges as \emph{``[they] are technically sound in creating maps but have limitations in terms of available [funding and personnel]. There is no open source tool like \app so what you have done is excellent.''}
\expert{1} enthusiastically said, \emph{``I'm [going to] mess around with this on my own, as this is really cool.''}
\expert{4} similarly said, \emph{``I don't know what I want to click on next. I just like being able to see so many [binning methods] all at once.''}

\paragraphHeadingSpace\bpstart{\emph{``My knowledge was enhanced today.''}--\expert{10}}

\noindent Among established binning methods, our experts used \emph{pretty breaks} and \emph{manual interval} most often followed by \emph{natural breaks}, \emph{standard deviation}, \emph{quantile}, and \emph{unclassed}. 
Many experts were unfamiliar with certain methods, e.g., \emph{ck-means} (\expert{4,15}), \emph{head-tail breaks} (\expert{3,4,15}), and \emph{box plot} (\expert{1}).
This variance is due to personal preferences (\emph{``I just use the ArcGIS default, natural breaks''}--\expert{16}), organizational policies (\emph{``it is a practice to use standard deviation here''}--\expert{1}), 
prior choices (\emph{``if there is a report that goes out every year, we also re-use the binning method every year, irrespective of the new data distribution''}--\expert{6}), 
and on the data (\emph{``equal interval is used for soft-range data, geometric or manual interval is used for long-range data''}--\expert{3}).

\paragraphHeadingSpace\bpstart{\emph{``It can surprise mapmakers, challenge their initial assumptions.''}-\expert{4}} 

\noindent Our experts suggested using \app to \emph{``promote thinking about how best [to] display the data''} (\expert{9}), \emph{``perform initial data exploration''} (\expert{1}), \emph{``reveal things [one hadn't] thought about''} (\expert{4}), as a verification tool \emph{``to just check something before [eventually] making the map in their own systems''} (\expert{7}), support \emph{``wireframing or prototyping along with ColorBrewer''} (\expert{13}), \emph{``to very quickly demonstrate to project managers the importance of class breaks''} (\expert{15}), and as a \emph{``quicker, interactive way for quality checking purposes.''} (\expert{13}).

\paragraphHeadingSpace\bpstart{\emph{``It could be dangerous in the hands of somebody who doesn't know how to bin data or is nefariously trying to skew the data.''}--\expert{8}} 

\noindent Four experts (\expert{2,4,8,15}) noted ways in which \app could be misused.
\expert{2} remarked, \emph{``I know this sounds bad but [the Paint Mode] lets me decide how much manipulation of data I want.''}
\expert{15} shared one of their clients' comment on a map draft: \emph{``It doesn't highlight the three districts that I need. Change those [bin extents] so that [the districts] show as bright red, and people think there is the problem''} and later noted how it defeated the purpose of making an unbiased map.
\expert{4} \emph{``[were] hopeful things don't get abused as it is very easy to mislead with graphics.''} 
On the contrary, two experts (\expert{5,14}) endorsed the \emph{Paint Mode}. 
\expert{14} found the ability to \emph{``paint counties with the same color as roughly similar, and everything in the bins above as higher, as very interesting to facilitate comparisons.''}
\expert{8} aptly summed it, \emph{``Maps are not perfect representations of reality but we do our best and present data in a way that maximizes the public good. [In spite of these concerns,] I think this is a really useful tool.''}

\subsection{View Specific Feedback}

\paragraphHeadingSpace\bpstart{Browse View.} 
\expert{9} acknowledged that they were \emph{``not used to such a large number of [binned] representations''} and that the \emph{Browse View} \emph{``promotes thinking about how best we can display the data.''}
\expert{2} found the short descriptions for each binning method useful. \expert{6} \emph{``liked the interactivity where you could do the roll over [on the map] and see what the values are in those individual counties [in a tooltip].''}

\paragraphHeadingSpace\bpstart{Compare View.} All experts found the 
the \emph{Compare View} to be useful. \expert{12} found \emph{``comparing the distributions [this way] very cool because it clearly quantifies what you can visually [see and] tell from the map, e.g., the bright green [bin] has way more counties in it than the other one.''}
\expert{5} said, \emph{``I am glad you are including this chart here. It is actually something we are considering adding to a map we are working on right now. I like it, I think it's very useful.''}

\paragraphHeadingSpace\bpstart{Combine View.} 
Observing or interacting with the \emph{resiliency} binning method~\citep{narechania2023resiliency} in the \emph{Combine View} sparked intriguing thoughts in our experts, including identification of potential use-cases. 
\expert{15} were intrigued, \emph{``as in a lot of ways this [resiliency] is the `best way' to display the data because it is suggesting the average of all the different bins that it could fall into, [i.e.,] the one that makes the most sense across different binning methods, so I like that.''}
\expert{8} wondered \emph{``Why didn't [they] think of this stuff, it seems so simple on the face of it;''} they found it a \emph{``really interesting binning method that one could use for `just about anything'.''}
\expert{3} reserved their feedback until after analyzing it in depth.

In terms of use-cases, \expert{14} suggested utilizing this view \emph{``if as a mapmaker you want to avoid having to make decisions about binning methods, especially in a situation where there is possibility of very technical criticism, you can just point to the resiliency algorithm and say, I just let that decide. As a government employee, I frequently have to justify decisions that I make, and I don't quite get to arbitrarily say, `Oh, I just wanted to do it this way'.''}
\expert{14} also identified \emph{``a flip side that the checkboxes [to specify which binning methods to use to compute resiliency] imply that user decision is still required [and that its not a fully automated technique].''}
\expert{1} noted another use case to see where counties are consistent across different binning methods and accordingly highlight the counties that we might be interested in.
\expert{6} suggested an interesting application to use \emph{resiliency} to combine the bins in annual reports across multiple years as an average.
\expert{2} had interesting observations while interacting with it in the \emph{Combine View}, e.g., \emph{``It does a better job at highlighting an outlier, by showing the areas with lower life expectancy as well''} and \emph{``quantile seems to have a big impact on resiliency.''}
\expert{13} identified it \emph{``as a way to clearly highlight information in public health''} but questioned \emph{``how digestible and understandable is it for a general audience?''}

\paragraphHeadingSpace\bpstart{Create View.} 
\expert{5} found the \emph{Create View} really useful \emph{``especially when you're trying to highlight certain features, e.g., while making maps for policymakers that are focused on specific area to be able to highlight certain values in their congressional district and wanting to bin two or more regions into the same bin. I think it's an interesting thought I haven't seen anywhere!''} 
\expert{14} said, \emph{``When you first showed that to me, I was like, oh, that's cool, but I don't know what I would do with it. The comparison it lets you do by [painting counties] with the same color as roughly similar, with everything in the bins above is higher is very interesting.''}

\subsection{Suggested Applications and Use Cases}
Our experts suggested multiple use cases for \app in education, journalism, and other business settings.

\paragraphHeadingSpace\bpstart{Create and export quick maps with limited tooling and expertise.}
Three experts noted \app's utility for domain novices to upload data, visualize it, create quick maps, and export them independently without expert assistance (\expert{6,7,12}). \expert{2} highlighted its value for users without GIS tools or agencies lacking GIS expertise but with JavaScript skills to utilize the source code export feature (\expert{6}). \expert{3} appreciated \app's customizability for developing open-source applications.

\paragraphHeadingSpace\bpstart{\emph{``It is like a professor's `dream tool'''}--\expert{4}}
 
\noindent Four experts suggested using \app as an educational resource to learn about unfamiliar binning methods~(\expert{1,4,9,14}; \emph{``We are not used to such a large number of [binning methods]''}--\expert{9}), teach about map design~(\expert{4}), and instruct how the choice(s) behind a binning method can make a map look very different and affect key reader takeaways~(\expert{14}).

\paragraphHeadingSpace\bpstart{It can help journalists, fund managers, policymakers, businesses.}
Our experts also suggested that \app can help journalists \emph{``to imagine different ways of telling a totally different story''} (\expert{4}), fund managers \emph{``to custom bin townships to qualify for funding opportunities''} (\expert{5}), policymakers \emph{``to highlight certain features by binning certain congressional districts together''} (\expert{8}), and  businesses \emph{``to inspect gross trends in transactions''} (\expert{2}).

\section{Takeaways for Mapmakers, Researchers, and Tool Developers}
\label{section:takeaways}
From our interviews, we derived the following set of takeaways for mapmakers as well as researchers and tool developers. Many of these corroborate cartography education and standards while others put these standards into perspective, informing future studies and tools.

\paragraphHeadingSpace\bpstart{Jenks \& Coulson's (\citeyear{jenks1963class}) Five Rules for Binning are generally still valid.} Users should:
(1) encompass the full range of the data; (2) have neither overlapping values nor vacant bins; (3) have enough bins that are warranted by the nature of the collected data without sacrificing its accuracy; (4) divide the data into reasonably equal observation groups; and (5) have a logical, mathematical (not arbitrary) relationship to data if it is practical.

\paragraphHeadingSpace\bpstart{Terminology may vary across users and organizations.} Throughout the interview study process, we noticed our experts interchangeably used ``choropleths'', ``chlorofills'', ``thematic maps'', ``colored maps'', and even just ``maps''. 
Even ``bins'', ``classes'', ``categories'', ``buckets'', ``schemas'', and ``ranges'' were used interchangeably. 
These terminologies can create confusion during recruitment because our invitees sometimes either pointed us to other experts or altogether declined the invitation claiming they did not make those maps (we knew they did!).

\paragraphHeadingSpace\bpstart{There is a need for more open-source tooling.} There is no single, ideal toolset that works; mapmakers often use a combination of tools depending on the size and nature of the underlying data, team dynamics in the work environment, intended use-case(s) and target audience, tool's characteristics, and own expertise and experience working with the tool(s). While larger organizations can afford proprietary Esri products, many mapmakers welcomed the open-source and customizable status of \app and advocated for more such solutions.

\section{Tool Limitations, Expert Requests, and Future Work}
\label{section:limitations}
With inputs from our experts, we noted certain limitations and identified several opportunities for future work.
First, \app currently supports sixteen binning methods; we plan to support more methods, especially from the \emph{spatial} (e.g., \emph{equal area}) and \emph{iterative} (e.g., \emph{DBSCAN}) categories.
Next, to better integrate our tool into existing mapmaking workflows, our experts requested supporting additional input file types -- such as shapefiles, KML, and KMZ (\expert{9}); and also additional output file types such as JPEG, (\expert{12}), PDF (\expert{12}), and shapefiles (\expert{9}).
\expert{3} requested the ability to programmatically specify custom binning methods either by uploading or typing a custom function.
\expert{7} requested mechanisms for comparing multiple variations of the same binning method, e.g., \emph{``I often like to look at multiple quantiles [like 3 4 5, and 6] next to each other, I think that would be super helpful.''}
We also received requests to enable resizing maps in relevant views (\expert{6}), allow searching and zooming into specific map regions (\expert{4,6}), and to provide video demonstrations of how to use the tool (\expert{2}).

In addition to the above expert feature requests, we have also identified several opportunities for future work. 
First, in the \emph{Browse View} and \emph{Compare View}, we plan to enable simultaneous comparisons of binning method outputs across different color schemes--such as divergent, sequential--to aid better map comparisons and exports.
We also plan to superimpose a cumulative frequency curve (as a line chart) on the existing bar charts (in the \emph{Compare View}), to provide an alternate visualization of the relation between bin breaks and corresponding bin extents~\citep{andrienko2004cumulative}.

Next, to help mapmakers create accessible choropleth maps, future work is planned to support hatching (or stippling)--a technique where the choropleth polygons are filled with patterns of lines, dots, or other marks instead of solid colors to help users differentiate between different categories or levels of data, especially for colorblind users.

Next, in the \emph{Create View}, we plan to elicit additional input from the user--such as the dataset's semantics and the map's target audience--to recommend the `best', `most appropriate' binning method based on these requirements.
Similarly, we also plan to process the user-created custom bins--through \emph{Manual~Configuration} and/or \emph{Paint~Mode}--and recommend relevant binning outputs of `closest' and `established' methods instead.
Next, in addition to the warning next to the \emph{Paint~Mode}, we also plan to link real-world examples of how this feature can be (but should not be) misused for propaganda purposes.
All of the above approaches are aimed at increasing awareness of and mitigating the potential harms of creating misleading maps.

Finally, we suggest incorporating binning methods into cutting-edge tests for map trustworthiness. Prior research has found that adding metadata, and using professional styles may enhance trust~\citep{prestby2025trust}; this could be extended to examining binning methods as well.

\section{Conclusion}
\label{section:conclusion}
Mapmakers often bin continuous data values into discrete bins when making choropleth maps. In this work, we present
\textbf{\app}, a new, open-source, web-based geospatial visualization tool that helps users browse a catalog of sixteen established binning methods and compare, customize, and export custom maps. This tool advances the state of the art by providing multiple binning methods in \emph{one view} and supporting administrative data reclassification \emph{on-the-fly}. 
Sixteen cartographers and GIS experts provided feedback on \app, and identified opportunities to integrate it into their mapmaking workflow. They also shared that it could be used to educate seasoned mapmakers and new mapmakers, and to illustrate the importance of choosing appropriate binning methods.
Exploropleth is open-source and publicly available at \textbf{\url{https://exploropleth.github.io}}.

\section*{Disclosure statement}
The authors do not have any relevant financial or non-financial competing interests to report.

\section*{Funding}
No funding was received for this work.

\section*{Data availability statement}
The two datasets--``Life Expectancy'' and ``Population Density'' of U.S. counties in 2019--used to demonstrate our tool and its use-cases are made openly available by the Institute for Health Metrics and Evaluation (\url{https://www.healthdata.org/}) and the U.S. Census Bureau (\url{https://www.census.gov/data.html}), respectively.

\bibliographystyle{apacite}
\bibliography{references}

\end{document}